\documentclass[fleqn,12pt]{wlscirep}
\usepackage[utf8]{inputenc}
\usepackage[T1]{fontenc}
\usepackage{etoolbox}
\DeclareUnicodeCharacter{2500}{?}

\newcommand{\nustar}{\textit{NuSTAR}\ }

\usepackage{comment}

\def\aa{Astron. Astrophys.}

\def\aap{A\&A}

\def\apj{Astrophys. J.}

\def\apss{Ap\&SS}
\def\apjl{Astrophys. J.}

\def\araa{ARA\&A}

\def\mnras{Mon. Not. R. Astron. Soc.}
\def\nat{Nature}
\def\pasp{PASP}
\def\pasj{PASJ}

\def\physrep{Phys.~Rep.}

\usepackage{lineno}

\usepackage{bm}
\DeclareCaptionLabelSeparator{mysep}{$\boldsymbol{|}$}
\title{\center A shared accretion instability for black holes and neutron stars}
 \author[1,2,3,4$\star$]{F. M. Vincentelli}\author[3]{J. Neilsen}
\author[5,$\dagger$]{A.J. Tetarenko}
\author[6,7]{Y. Cavecchi}
\author[4]{N. Castro Segura}
\author[8]{S.~del~Palacio}
\author[9]{J.~van~den~Eijnden}
\author[10,11]{G.~Vasilopoulos}
\author[4]{D.~Altamirano}
\author[1,2]{M.~Armas~Padilla}
\author[10]{C.~D.~Bailyn}
\author[12]{T.~Belloni}
\author[4]{D.~J.~K.~Buisson}
\author[1,2]{V.~A.~C\'{u}neo}
\author[13]{N.~Degenaar}
\author[4]{C. Knigge}
\author[14,15]{K. S. Long}
\author[1,2]{F. Jim\'enez-Ibarra}
\author[16]{J.~Milburn}
\author[1,2]{T.~Mu\~{n}oz~Darias}
\author[4]{M. Ozbey Arabaci}
\author[17]{R. Remillard}
\author[18]{T. Russell}

\affil[1]{Instituto de Astrof\'{i}sica de Canarias, E-38205 La Laguna, Tenerife, Spain}
\affil[2]{Departamento de Astrof\'{ı}sica, Universidad de La Laguna, E-38206 La Laguna, Tenerife, Spain}
\affil[3]{Villanova University, Department of Physics, Villanova, PA 19085, USA}
\affil[4]{Department of Physics and Astronomy, University of Southampton, SO17 1BJ, UK}
\affil[5]{Department of Physics and Astronomy, Texas Tech University, Lubbock, TX 79409-1051, USA}
\affil[6]{Instituto de Astronom\'ia, Universidad Nacional Aut\'onoma de M\'exico, Ciudad de M\'exico, CDMX 04510, Mexico}
\affil[7]{Departament de F\'sica, EEBE, Universitat Polit\`ecnica de Catalunya, Av. Eduard Maristany 16, 08019 Barcelona, Spain }
\affil[8]{Department of Space, Earth and Environment, Chalmers University of Technology, SE-412 96 Gothenburg, Sweden}
\affil[9]{Astrophysics, Department of Physics, University of Oxford, Denys Wilkinson Building, Keble Road, Oxford OX1 3RH, UK}
\affil[10]{Department of Astronomy, Yale University, PO Box 208101, New Haven, CT 06520-8101, USA}
\affil[11]{Universit\'{e} de Strasbourg, CNRS, Observatoire astronomique de Strasbourg, UMR 7550, F-67000 Strasbourg,
France}
\affil[12]{INAF - Osservatorio Astronomico di Brera, Via E. Bianchi 46, I-23807 Merate, Italy}
\affil[13]{Anton Pannekoek Institute for Astronomy, University of Amsterdam, Science Park 904, 1098 XH Amsterdam, The
Netherlands}
\affil[14]{Space Telescope Science Institute, 3700 San Martin Drive, Baltimore, MD 21218, USA}
\affil[15]{Eureka Scientific, Inc. 2452 Delmer Street, Suite 100, Oakland, CA 94602-3017, USA}
\affil[16]{Cahill Center for Astronomy and Astrophysics, California Institute of Technology, Pasadena CA 91125, USA}
\affil[17]{MIT Kavli Institute for Astrophysics \& Space Research, MIT, 70 Vassar Street, Cambridge, MA 02139, USA}
\affil[18]{INAF, Istituto di Astrofisica Spaziale e Fisica Cosmica, Via U. La Malfa 153, I-90146 Palermo, Italy}
\affil[$\dagger$]{NASA Einstein Fellow}

\begin{document}
\flushbottom
\maketitle

{\bf
\noindent
Accretion disks around compact objects are expected to enter an unstable phase at high luminosity \cite{lightman1974}. {One instability may occur when} the radiation pressure generated by accretion modifies the disk viscosity, resulting in the cyclic depletion and refilling of the inner disk on short timescales \cite{belloni1997}. Such a scenario, however, has only been quantitatively verified for a single stellar-mass black hole \cite{janiuk2000,nayakshin2000,neilsen2011}. {Although there are hints of these cycles in a few isolated cases}\cite{bagnoli2015,altamirano2011,janiuk2011,janiuk2015,kimura2016}, their apparent absence in the variable emission of most bright accreting neutron stars and black holes has been a lingering puzzle\cite{done2004}. Here we report the presence of {the same multiwavelength instability} around an accreting neutron star. 
Moreover, we show that the variability across the electromagnetic spectrum---from radio to X-ray---of both black holes and neutron stars at high accretion rates can be explained consistently if the accretion disks {are} unstable, producing relativistic ejections during transitions that deplete or refill the inner disk. 
Such new association allows us to identify the main physical components responsible for the fast multiwavelength variability of highly accreting compact objects.

}
\vspace{30pt}

Swift J1858.6$-$0814 (hereafter Swift J1858) is a low mass X-ray binary (LMXB) that was first detected in November 2018\cite{krimm2018}  and reached a maximum X-ray luminosity of $\approx10^{37}$ erg s$^{-1}$ (0.6--79 keV)\cite{hare2020}. Spectral analysis showed peculiar properties, including significant  obscuration\cite{hare2020,Koljonen2020} { (N$_H \approx10^{23}$~cm$^{-2}$)} and outflows in X-rays\cite{buisson2020_1858_wind}, optical\cite{munios-darias_2020} and UV\cite{castro-segura_2022}. Moreover, for more than a year after its discovery, the source showed remarkable flaring activity from radio to hard X-rays\cite{hare2020,vandeneijnden2020,buisson2020_1858_wind,buisson2020_1858_eclipse}. The source returned to quiescence in 2020, but not before exhibiting X-ray eclipses\cite{buisson2020_1858_eclipse} and Type-I X-ray bursts\cite{buison_2020_1858_burst} indicating the presence of an accreting neutron star with an orbital inclination >70$^\circ$ {at a distance of $\approx$13 kpc}. 
\smallskip

\begin{figure}
\centering
\includegraphics[width=1\columnwidth]{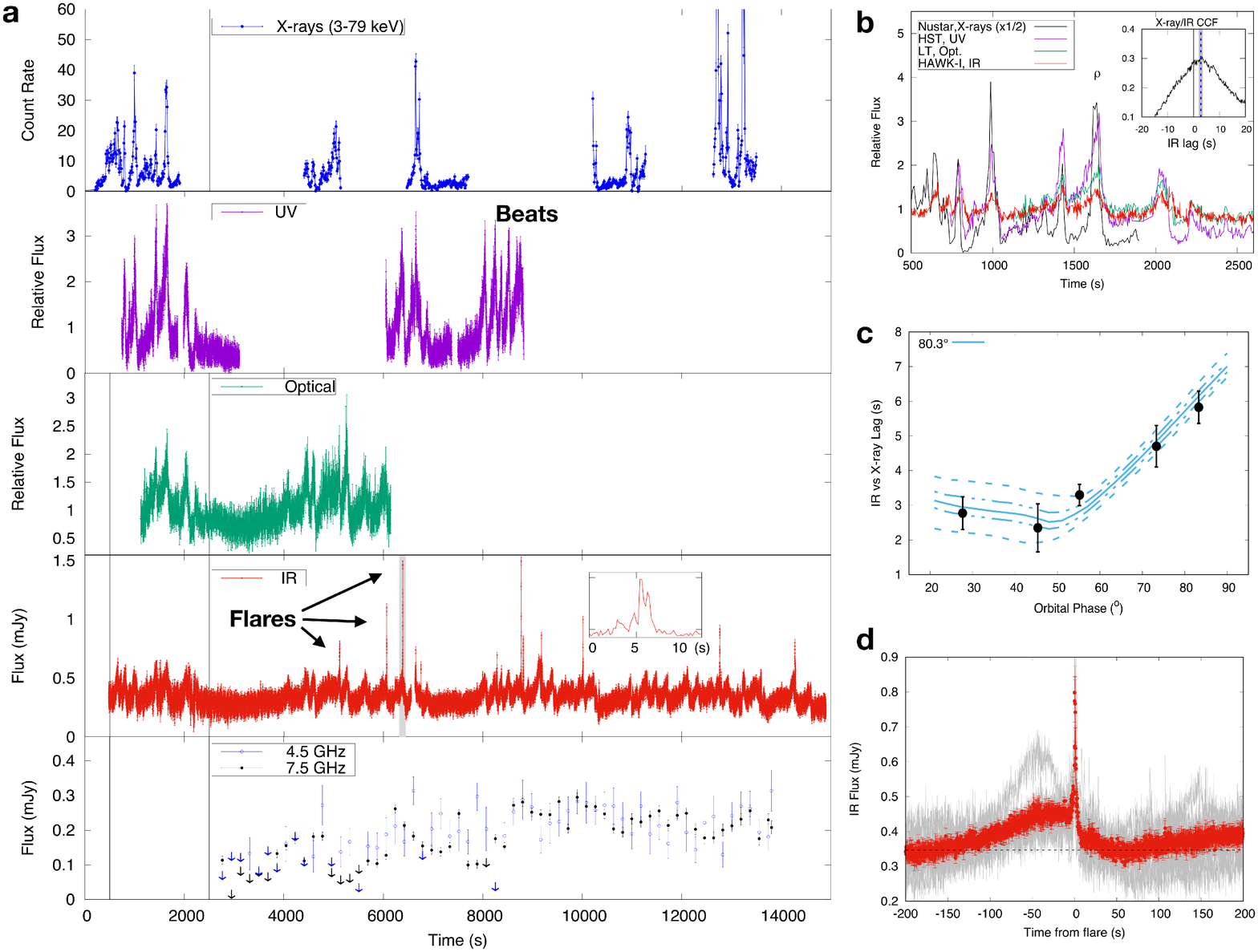}

\caption{ \textbf{Multiwavelength variability of Swift J1858 } (a)   Campaign of Swift J1858  with time resolved observations in (starting from the lowest panel) radio (VLA), IR (HAWK-I), optical (LT), UV (HST) and X-rays (\nustar). The zero-time is 06/08/2019 00:00:00 UTC. All lightcurves are plotted with their standard error. For the top four lightcurves an interpolation line is also plotted for clarity. The inset in the IR panel shows an example of a fast flare from the grey shaded area. (b) Strictly simultaneous X-ray UV/O/IR segment using rebinned lightcurves: all bands follow the same repetitive pattern. The inset shows the X-ray/IR CCF with a peak at 2.8$\pm$0.5 s (see Methods). (c) X-ray vs IR lag as a function of the orbital phase of Swift J1858. Errors are computed as the standard deviation of the lag distibution (see Methods). The solid line represents the expected best fitted delays from an irradiated disk and companion star in a neutron star LMXB with an orbital period of 21.3~h  and an inclination of $\approx$80$^\circ$. Dot-dashed and dashed lines corresponds to the 68\% and 99\% confidence intervals. (d) The average lightcurve profile of the fast IR flares over 400 s. Errorbars show the standard error. The grey curves show sovrapposition of the individual lightcurves used for the stack.}
\label{fig:dataset}
\end{figure}

\begin{figure}
\centering

\includegraphics[width=1.0\columnwidth]{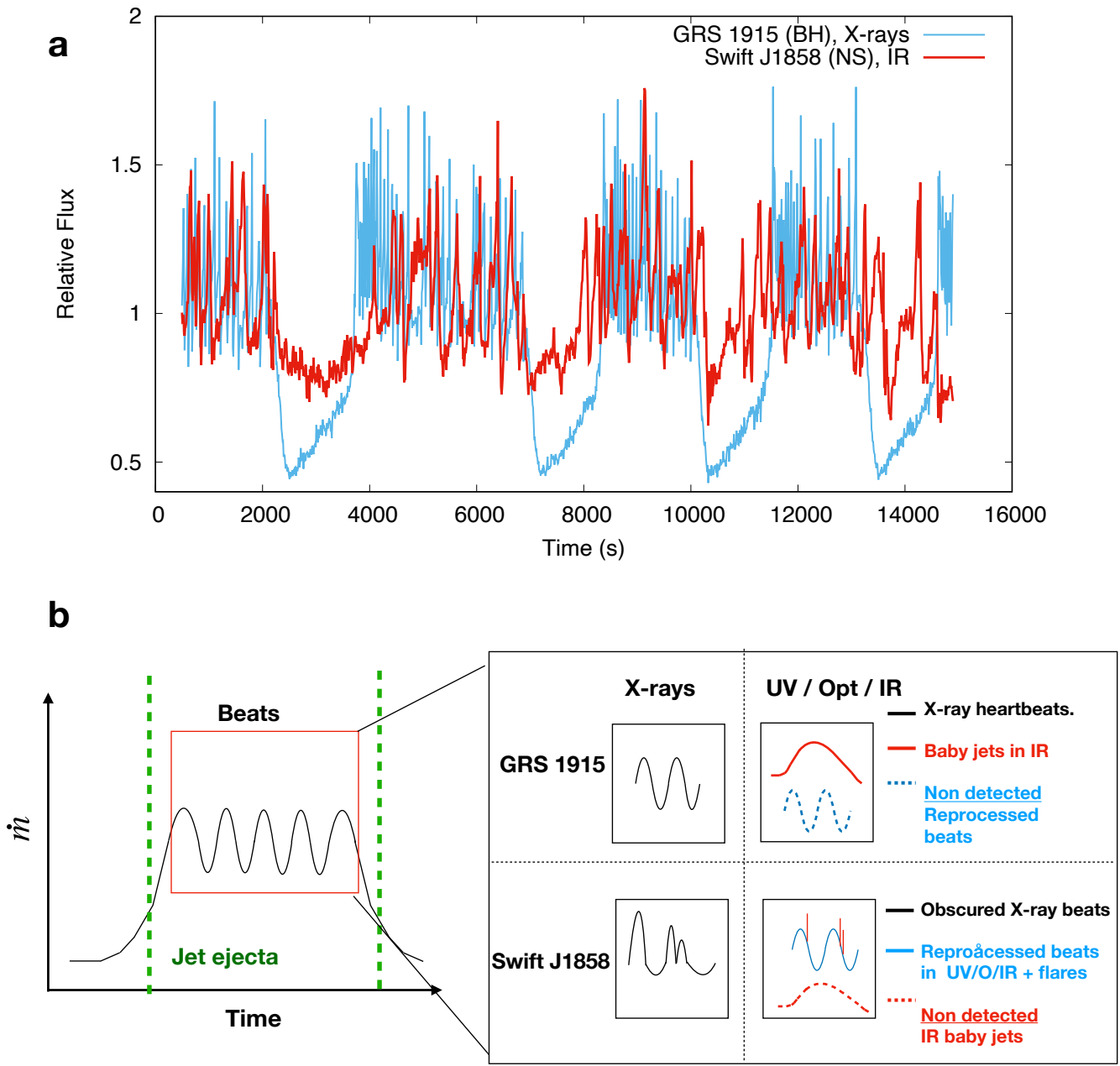}
\caption{  \textbf{The unified accretion instability for black holes and neutron stars}  (a) HAWK-I lightcurve of the accreting neutron star Swift J1858 (red line) compared with the \textit{Chandra} X-ray lightcurve of the accreting black hole GRS 1915+105 (blue line). The lightcurves were rebinned to 10 s and aligned to their minimum value (at $\approx$11000 s). The similarity of the limit cycle timescales is evident and strongly indicates a common physical mechanism.   (b) Cartoon of the underlying physical mechanism that explains the multiwavelength phenomenology of both sources. A limit cycle in the accretion rate $\dot{m}$ generates ``beats" (or ``heartbeats") when the disk reaches the luminous phase, along with radio ejecta (``baby jets"). Fast IR flares, instead, originate from optically thin synchrotron emission, which does not contribute to the overall radio emission. Depending on the contribution from obscuration, reprocessing, and the jet, the beat oscillations can be detected in X-rays or at UV/O/IR wavelengths.}
\label{fig:comparison}
\end{figure}

On the 6th of August 2019, we coordinated a multiwavelength campaign to observe Swift J1858 simultaneously for $\sim$4 h with high time resolution in 5 bands: X-rays (3--79 keV) with \nustar; UV (150 nm) with the {\em Cosmic Origins Spectrograph} onboard the Hubble Space Telescope; optical ({\it i}+{\it z sdss} band, effective wavelength $\lambda_\mathrm{eff} =720$ nm) with the \textit{RISE} at the Liverpool Telescope; near-IR ($K_s$ band, $\lambda_\mathrm{eff} = 2.2~\mu$m) with HAWK-I on the Very Large Telescope; and radio ($4.5$ and $7.5~{\rm GHz}$) with the Karl G. Jansky Very Large Array. The source showed very strong variability with similar patterns in UV, optical, IR (UV/O/IR), and X-ray (see Figure ~\ref{fig:dataset}-a-b). On long timescales, Swift J1858 exhibited a repetitive behaviour, alternating between quiet and active/variable phases    (Figure~\ref{fig:dataset} and Figure~\ref{fig:comparison}). The active phases showed oscillatory behavior on timescales of $\approx$100 s; we refer to these as "beats," given their visual similarity to the "heartbeat" variability pattern in GRS 1915+105 \cite{neilsen2011}. On timescales of seconds, the source showed episodic fast flaring events (seen only in IR), which we refer to as "flares".
\smallskip

To explore the multiwavelength temporal behavior, we computed the cross-correlation function (CCF) between \nustar and HAWK-I for all the simultaneous segments in our dataset (see Methods). We measured a clear correlation between the two bands, but the IR lags the X-ray variability with a delay that changes from $\approx 2.5$~s to $\approx5.5$~s (see Figure~\ref{fig:dataset}-c). The magnitude and orbital phase dependence of these lags are fully consistent with a model where the UV/O/IR beats originate from the irradiation of X-ray beats on a disk and donor star with high orbital inclination ($\approx80^\circ$) and the orbital period of Swift J1858 ($\approx$21.3 h\cite{buisson2020_1858_eclipse}). 
\smallskip

Simple mass accretion rate variations in a hot inflow are not likely to explain the driving X-ray {lightcurve}\cite{belloni1997}. The X-ray variability observed in Swift J1858 shows significant spectral evolution not compatible with standard variability of accreting compact objects \cite{belloni2000,janiuk2000,nayakshin2000}. In addition, similar variability has been seen in the archetypal high accretion rate stellar-mass black holes GRS 1915+105 and  V404 Cyg \cite{hare2020}. {These sources also share other important properties with Swift J1858, such as high luminosity (40\% of the Eddington luminosity for Swift J1858), obscuration and ouflows}\cite{hare2020,Koljonen2020}. 
This association is strengthened by the remarkable similarity of the IR lightcurve of Swift J1858 and the X-ray lightcurve of the so-called "$\beta$" variability class of GRS 1915+105\cite{belloni2000} (Figure~\ref{fig:comparison}). Even though the patterns are less discernible in the X-ray band for Swift J1858 (probably due to variable line-of-sight obscuration, given its high inclination\cite{janiuk2015,hare2020,buisson2020_1858_wind,munios-darias_2020}), the irradiation origin of the UV/O/IR lightcurve strongly suggests a common physical mechanism for the driving variability in both sources. 
\smallskip

From a physical point of view, it is commonly accepted that the recurrent behaviour of GRS 1915+105 (i.e., heartbeats and other limit cycles) is due to a radiation pressure instability in the disk at high accretion rates \cite{belloni1997,janiuk2000,nayakshin2000,neilsen2011}.  {Although not fully confirmed by GRMHD simulations, this instability is believed to drive cyclic accretion or ejection and rebuilding of the inner disk}, generating repeating patterns in X-rays on 10--1000~s timescales \cite{janiuk2000,nayakshin2000,neilsen2011}. If this emission irradiates the disk and companion star, it will give rise to a delayed UV/O/IR lightcurve, such as the one observed in Swift J1858. {The interpretation of beats as a disk instability can be  tested:  both models\cite{nayakshin2000} and observations\cite{neilsen2011} of GRS 1915+105 need short-lived jet ejections near the peak luminosity (roughly coincident with the depletion of the disk). }

The fast IR flares in Swift J1858 appear to verify this hypothesis, {giving credence to the radiation pressure instability interpretation of the limit cycles}. Aligning and averaging the flares, including 200 s of data before and after each flare, reveals that they take place after the peak of the slower IR beats (see Figure~\ref{fig:dataset}-d). But these flares are inconsistent with a thermal origin (see Methods), and, given their red color,  we interpret them as direct evidence of optically-thin synchrotron emission from transient/short-lived relativistic jet ejections expected to occur\cite{nayakshin2000} during these beat oscillations. 

\smallskip

\begin{figure} 
\centering

\includegraphics[width=\columnwidth]{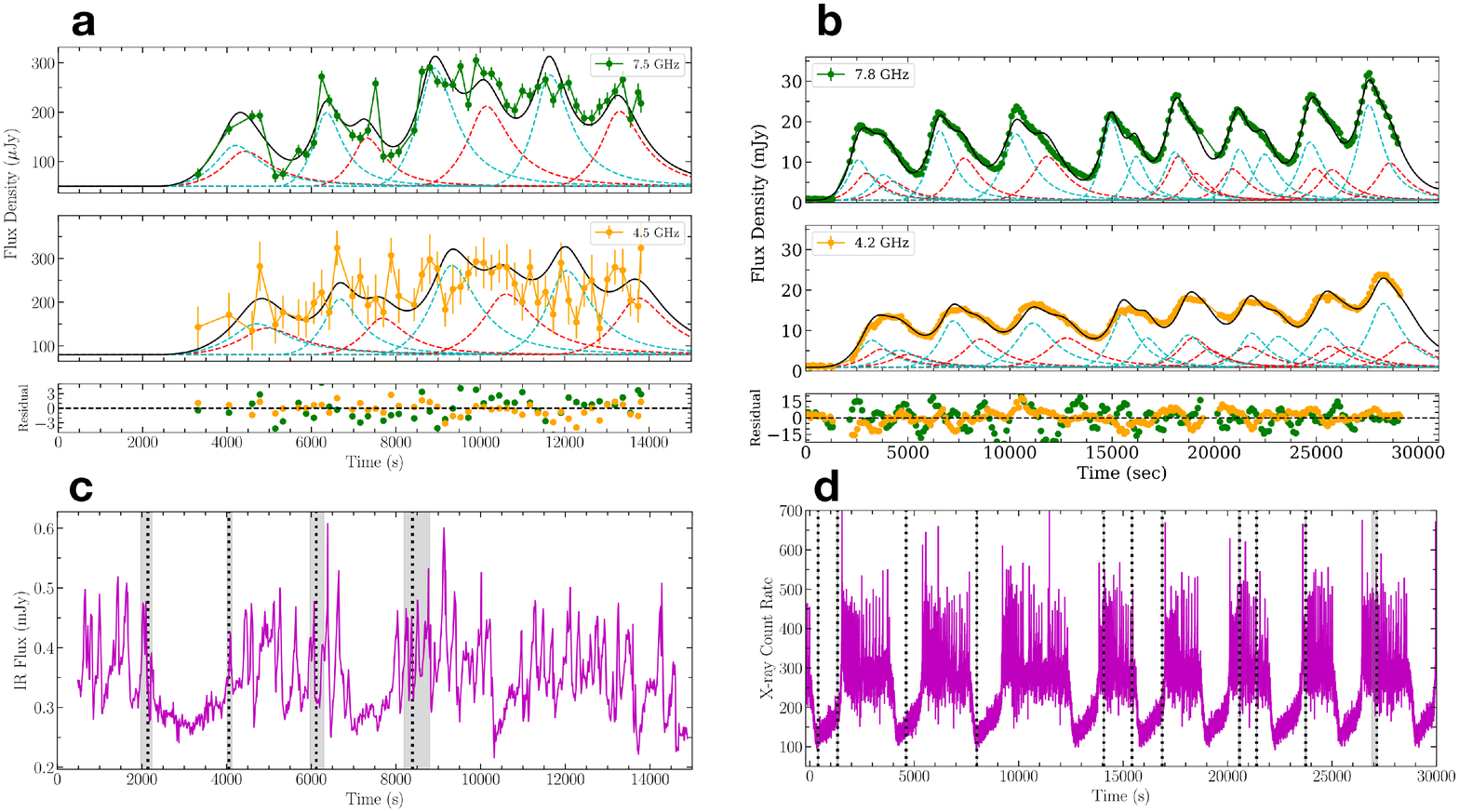}
\caption{\textbf{Radio oscillations modelling}  (a,b)  Radio time series in the two frequency bands (green and yellow points), over-plotted with the total jet model (solid black line) and the individual model contributions from the jet and the counter-jet for each ejection (blue and red dashed lines) of Swift J1858 (a) and GRS 1915(b). Residuals (data-model/uncertainties) are also shown for each fit. (c,d) Modeled jet ejection times (black dashed lines, grey shading indicates 1$\sigma$ confidence intervals) with respect to the IR for Swift J1858 (c) and X-ray for GRS 1915 (d). }
\label{fig:radio-modelling}
\end{figure}

Swift J1858 also showed significant radio variability throughout our campaign\cite{vandeneijnden2020}, which requires explanation. {The fast IR flares cannot be responsible for the observed low-frequency variability because their amplitude and duration would naturally lead to their radio emission being completely self-absorbed ($\tau \gg 1$ at 10 GHz; see Methods).} 
However, observations of GRS 1915+105 also show "baby jets": strong radio flares (though their synchrotron emission can contribute significantly in the IR band\cite{fender1998,eikenberry1998}) that are consistent with emission from adiabatically expanding blobs\cite{mirabel1998} {(although their launching mechanism is still not clear)}. To search for baby jets in Swift J1858 and make a comparison to GRS 1915+105, we modeled its variable radio emission as the sum of multiple ejecta\cite{tetarenko2017}, performing the same analysis on an archival radio observation of GRS 1915+105 (coincident with the $\beta$-class X-ray lightcurve shown in Figure~\ref{fig:comparison}). The results presented in Figure~\ref{fig:radio-modelling} show that the radio variability of both sources is well reproduced by our modelling. For Swift J1858, the model suggests baby jet ejection times (grey shaded areas in Figure~\ref{fig:radio-modelling}) near quiet/active phase transitions; most of the ejecta in GRS 1915+105 occur during quiet phases but several fall close to quiet/active transitions as well.

\smallskip

 For self-consistency, we then tested whether Swift J1858's baby jets would be detectable in the IR as for GRS 1915+105. Past studies\cite{mirabel1998,neilsen2011} show accretion instabilities in GRS 1915+105 when the X-ray and radio luminosity are $L_\mathrm{BH_X}\approx 10^{38}$~erg\,s$^{-1}$ and $L_\mathrm{BH_{radio}}\approx 10^{30}$~erg\,s$^{-1}$, respectively. For Swift J1858, we find $L_\mathrm{NS_X}\approx10^{37}$~erg\,s$^{-1}$ and $L_\mathrm{NS_{radio}}\approx10^{29}$~erg\,s$^{-1}$ \cite{vandeneijnden2020}. 
Even under the conservative assumption that the ratio between the IR and radio flux from the jet in Swift J1858 is the same as the one observed in GRS 1915+105 during the $\beta$-class instability (IR/radio $\approx$ 1.4)\cite{mirabel1998}, then we expect an IR baby jet flux of only $\approx$0.24 mJy. This is almost a factor of two fainter than the reprocessed emission during the beats ($\approx$0.4 mJy). This indicates that the two sources share the same disk-jet coupling, despite having qualitatively different radio and IR lightcurves.
 More broadly, {regardless of the jet launching mechanism}, this shows how the appearance of accretion instabilities can depend not only on the accretion rate and disk-jet geometry, but also on the binary orbit and the mass of the compact object.

\smallskip
 There is growing evidence that high-accretion rate black hole sources such as GRS 1915+105, V4641~Sgr, Cyg X-3, and V404 Cygni all share common X-ray spectral variability properties\cite{Koljonen2020}. However, multiwavelength parallels have proven more difficult due to their different extinctions, hampering efforts to produce a unified physical scenario for this class of sources. Yet, { as envisioned from our conclusions}, Swift J1858  shows clear analogies with all these objects. Simultaneous multiwavelength observations of the 2015 outburst of V404 Cygni revealed repetitive optical/X-ray patterns with a lag consistent with reprocessing\cite{kimura2016,alfonso2018,hynes2019} and {fast non-thermal flares}\cite{2017Sci...358.1299D}. 
Furthermore, its extreme radio variability is consistent with jet ejections taking place \emph{during X-ray spectral transitions}\cite{tetarenko2017}. Moreover, similar O-IR recurrent patterns with comparable timescales have also been observed in V4641 Sgr\cite{uemura2004} and Cyg X-3\cite{fender1996}. Finally, we note that X-ray heartbeats have also been detected in sources like the LMXB IGR J17091$-$3624\cite{altamirano2011} and the ULX NGC 3261\cite{motta2020}, which also shows significant line-of-sight obscuration despite having a lower inclination.{ Thus, the recent association of Swift J1858 as a low-luminosity Z-source\cite{Rhodes2022}, and the isolated presence of X-ray "GRS 1915-like" patterns in other accreting NSs such as the Rapid Burster\cite{bagnoli2015} and the Bursting Pulsar\cite{court2018}, strongly indicate that Swift J1858 represents the missing link for multiwavelength variability in high accretion rate sources (Figure~\ref{fig:comparison}, and Extended Data Figure \ref{fig:lvsb}).}
\smallskip

{It was also noted during review that while the limit cycle timescale is similar in GRS 1915+105 and Swift J1858 (despite their very different masses; see Methods), the beat timescale is much shorter around the black hole in the example lightcurves shown in Figure \ref{fig:comparison}. In fact, GRS 1915+105 exhibits a wide range of beat durations in similar limit cycles\cite{belloni2000}, which suggests that the beats may represent a second instability timescale\cite{nayakshin2000} or may be affected by other factors in the accretion flow. One possibility is the jet power, which is expected to have a significant impact on the disk structure, and thus on the observed X-ray lightcurve\cite{nayakshin2000,janiuk2000}. A careful comparison of the time-dependent radio/O-IR properties in states or sources with different beat timescales\cite{rothstein2005} could further elucidate the role of jets in shaping these instabilities. 
}

Our results draw a new coherent picture that links together key aspects of the multiwavelength variability of both black holes and neutron stars at high accretion rate: recurrent repetitive patterns, {radio oscillations} and fast flaring. At these high accretion rates, the accretion disk becomes unstable, resulting in disk-jet cycles on timescales of $\sim10$~s to $\sim1000$~s. These have historically been observed in X-rays, but our work shows that given the right conditions (e.g., inclination, orbital period, obscuration, and the relative brightness of the jet), accretion instabilities may in fact be more readily observable at UV/O/IR wavelengths. These instabilities are also 
observationally associated with radio-emitting discrete ejections: therefore, for the first time we can define a consistent physical scenario which can \emph{quantitatively} account for most of the multiwavelength variability observed from accreting compact objects at high luminosity. We argue that accretion instabilities, irradiation/obscuration, and jet ejecta should be seen as three fundamental pillars that can be used to study other classes of objects accreting near the Eddington limit.
With this insight, future time-resolved multiwavelength campaigns on compact objects will lead to better constraints on the physics of these instabilities and their hosts, independently of the nature of the central object\cite{janiuk2011}.

\newpage

\newpage
 
\subsection*{Acknowledgments}

 The authors thank the referees for the constructive comments which improved the manuscript. The interpretation of the  FV thank  R. Arcodia, P. Casella, G. Marcel, G. Mastroserio, N. Scepi and L. Stella for insightful discussions. The interpretation of the results benefited from discussions held during the meeting ‘Looking at the disc-jet coupling from different angles’ held at the International Space Science Institute in Bern, Switzerland. FV was supported by the NASA awards 80NSSC19K1456, 80NSSC21K0526 and from grant FJC2020-043334-I financed by MCIN/AEI/10.13039/501100011033 and NextGenerationEU/PRTR.  JN acknowledges support by the SAO award GO1-22036X. 
 AJT acknowledges support for this work was provided by NASA through the NASA Hubble Fellowship grant \#HST--HF2--51494.001 awarded by the Space Telescope Science Institute, which is operated by the Association of Universities for Research in Astronomy, Inc., for NASA, under contract NAS5--26555. D.A. and N.C.S. acknowledges support from the Science and Technology Facilities Council (STFC) grant ST/V001000/1.  FV, MAP and VC acknowledge support from the Spanish Ministry of Science and Innovation research project PID2020-120323GB-I00. MAP acknowledges support from the Consejería de Economía, Conocimiento y Empleo del Gobierno de Canarias and the European Regional Development Fund (ERDF) under grant with reference ProID2021010132 ACCISI/FEDER, UE. TMB acknowledges financial contribution from the agreement ASI-INAF n.2017- 14-H.0 and from PRIN-INAF 2019 N.15. TMD acknowledges support from the Spanish Ministry of Science and Innovation project PID2021-124879NB-I00,  and the Europa Excelencia grant (EUR2021-122010). TDR acknowledge financial contribution from the agreement ASI-INAF n.2017-14-H.0. 

\subsection*{Author contributions statement} FV and JN drew the new physical scenario for multiwavelength instabilities. FV carried out the multi-$\lambda$ timing analysis of the Swift J1858 dataset.  AJT modelled the radio variability with the help of SdP and JvdE. YC modelled the phase dependence of the X-ray/IR lag with the help of NCS.  For Swift J1858, NCS provided the the UV and X-ray data; FJI provided the optical data from LT; GV, CB JM and MOA provided optical data from Chimera and WASP; JvdE and TR provided the radio data. JN is the PI of the Chandra and VLA proposal on GRS 1915; AJT reduced and analysed the radio data. FV,  YC, GV, DA, TB,  ND, TMD, JvdE contributed significantly to the development of the luminosity-magnetic field  diagram (Extended Data Figure 1) to compare the different NSs. All authors contributed actively to the discussion and to the final version of the manuscript.

\subsection*{Competing Interest Statement}

The authors declare no competing interests

\newpage
\section*{Methods}

\subsection*{Data reduction: Swift J1858}

~~~~~~\underline{{\nustar}}

The Nuclear Spectroscopic Telescope Array (\nustar, 3--79 keV)\cite{harrison2013} observed the source between 2019-08-05T18:46:48 and 2019-08-06T08:26:45 UT (\textsc{obsid: 90501333002}). Barycenter-corrected events were extracted using the software \textsc{nupipeline}. Then, using \textsc{xselect} we filtered the events for both the FPMA and FPMB between 3 and 79 keV (i.e. between channels 35 and 2010) and within a region of 120" from the center of the source. We used the software \textsc{hendrics}\footnote{\url{http://ascl.net/1805.019}} to create the lightcurve with 7.8125 ms.

\underline{\textit{Hubble Space Telescope (HST)}}

Swift J1858 was observed in the far-UV with the {\em Cosmic Origin Spectrograph} and the G140L grating using the primary science aperture (PSA) with a total exposure of 4.9 ks (program GO/DD 15984, P.I. N. Castro Segura). The observations were obtained in TIME-TAG mode, recording events with a time resolution of $32\, \mathrm{ms}$. Data were reduced using the the HST {\sc calcos} and {\sc calstis} pipelines\footnote{Provided by The Space Telescope Science Institute (\url{https://github.com/spacetelescope})}.We extracted light curves from the {\sc time-tag} event files with a time resolution of  $32\, \mathrm{ms}$, using the same regions defined by the pipeline. An empirical background correction was applied. Regions affected by geocoronal airglow emission associated with $Lyman~\alpha$ ($\lambda 1208-1225$~{\AA}) and O~{\sc ii} ($\lambda 1298-1312$~{\AA}) were masked when extracting the light curves.

\underline{\textit{Liverpool Telescope}}

We obtained $4.5\, {\rm ks}$ ($1.25\, {\rm h}$) of high time-resolution photometry with the {\it RISE} fast-readout camera attached to the 2~m {\it Liverpool Telescope} at the {\it Observatorio del Roque de los Muchachos} during the first section of the multiwavelength campaign. The observations were carried out using the $720~{\rm nm}$ long pass filter ($\sim$ \textit{i+z} band). The fast readout time of RISE ($\sim 0.04$~s) resulted in 10~s time resolution. We used the data reduction pipeline for the optical imaging component of the Infrared-Optical suite. The flux calibration against a field star, catalogued in PanSTARRS (broad-band i filter), was performed using {\sc astropy-photutils}.

\underline{\textit{HAWK-I@VLT}}

We collected 4\,h of near IR time-resolved photometry in $K_s$ band (effective wavelength, $\lambda_\mathrm{eff} = 2.2~\mu$m) with {\it HAWK-I} \cite{pirard} mounted on the Very Large Telescope UT-4 {Yepun} under Program 103.201A.001. The instrument consists of four Hawaii 2RG $2048 \times 2048$ pixel detectors. However, by using the {\sc Fast-phot} mode, the instrument reads only a $128\times 64$ pixel window per quadrant, allowing us to achieve a time resolution of 0.2~s. Resulting images were stacked in ``data-cubes'' of 250 frames, separated by a readout gap of $\simeq 20\,{\rm s}$. The data were reduced with the ULTRACAM software tools \cite{dhilon2007}. Aperture parameters were derived using a bright reference star ($12.51~{\rm mag}$, $\sim 22$" west from the target) average point spread function and position during each cube. In order to mitigate the seeing effects the target and a comparison star ($13.11\pm 0.03$~mag, $\sim 33$" west from the target) were normalised to the reference star. Time stamps were put in a barycentered reference frame using the software developed by Eastman et al.\cite{eastman2010}.

\underline{\textit{VLA}}

Coordinated radio frequency observations of Swift J1858 were gathered with the {\it Karl G. Jansky Very Large Array} (VLA) on 2019 August 06 for $3.05$~h ($\sim {\rm 11~ks}$) under the program 19A--495 (P.I. van den Eijnden). The array was in the A-configuration, observing in C--band (4--8 GHz) with the 8-bit samplers. The correlator was setup to generate two base-bands centred at $4.5$ and $7.5~{\rm GHz}$, with a bandwidth of $1024~{\rm MHz}$ per base-band. For the observations, 3C286 was used as a flux calibrator and J1832$-$1035 was used as a phase calibrator. Flagging and calibration of the data were carried out within the Common Astronomy Software Application package (\textsc{casa} v4.7.2\cite{mc07}), with further details on the data reduction reported in the first radio study on this source\cite{vandeneijnden2020}.
To obtain a high time resolution light curve of the target source, we split the observation into a set of 3-minute timescale chunks and imaged each chunk in both base-bands. We used Briggs weighting with the robust parameter set to 0, to maximize sensitivity and reduce the side-lobe effects of a nearby background source during this imaging process. Flux densities were then measured by fitting a point source in the image plane (using the \texttt{imfit} task) for each time chunk. 

\subsection*{Data reduction: GRS 1915+105}
~~~~~~\underline{\textit{Chandra}}

\textit{Chandra} observed GRS 1915+105 for 120~ks beginning on 2011 June 21 at 04:36:42 UT. The data were taken with the High-Energy Transmission Grating Spectrometer (HETGS) in Graded Continuous Clocking (CC) mode, resulting in a 1-dimensional image with 8.55ms 
time resolution. The data were processed with standard {\tt ciao} data reduction tools, including {\tt acis\_process\_events}. We used {\tt axbary} to correct the arrival times of photons to the solar system barycenter. Because the source was so bright, telemetry saturation caused $\sim5$\% of the data frames to be lost from the detector chip ACIS-S3. We therefore used {\tt dmextract} to create 1-s lightcurves from each individual detector chip, then coadded them and interpolated over dropped frames to create the lightcurve in Figure~\ref{fig:comparison}.

\underline{\textit{VLA}}

Coordinated radio frequency observations of GRS 1915+105 were gathered with the VLA on 2011 June 22 for $10.8$~h ($\sim {\rm 39~ks}$) under the program SC0666 (P.I. Neilsen). The array was in the A-configuration, observing in the C--band (4--8~GHz). These data were taken in an early science mode where the correlator was setup to generate two base-bands centred at $4.2$ and $7.8~{\rm GHz}$, with a bandwidth of $128~{\rm MHz}$ per base-band. For these observations, 3C286 was used as a flux calibrator and J1856+0610 was used as a phase calibrator. Flagging and calibration of the data were carried out within \textsc{casa} v5.6, using standard procedures outlined in the \textsc{casa }Guides\footnote{\url{https://casaguides.nrao.edu}} for VLA data reduction (i.e., a priori flagging, setting the flux density scale, initial phase calibration, solving for antenna-based delays, bandpass calibration, gain calibration, scaling the amplitude gains, and final target flagging). To obtain a high time resolution light curve of the target source, we split the observation into a set of 2-minute timescale chunks and imaged each chunk in both base-bands. We used natural weighting to maximize sensitivity during this imaging process. Flux densities were then measured by fitting a point source in the image plane (using the \texttt{imfit} task) for each chunk.

\subsection*{Data analysis:}

\underline{\textit{Cross-correlation function: }}

We measured the cross-correlation function (CCF) between the \nustar and HAWK-I with a 0.2~s bin and a maximum shift of 100~s for each \nustar orbit. The very strong spikes in the IR can introduce spurious peaks in the correlation coefficient \cite{alfonso2018,hynes2019}$^,$\cite{gandhi2016,baglio2018}. Therefore we computed the CCF excluding and including the short IR flares (i.e. removing data points within $\pm1$~s of the flares).
We used the discrete CCF method\cite{edelson1988,gandhi2010}, rebinning the X-ray events to the HAWK-I time resolution. The lag was evaluated by simulating $N=10^4$ CCFs, randomizing each point using a Gaussian distribution with a $\sigma$ corresponding to the empirical error. For each run, the centroid was then computed with a weighted average of the correlation coefficient over a range of 80\% from the maximum\cite{gandhi2016}. The lag was then evaluated from the resulting distribution of centroid lags. The error was obtained from the standard deviation of the obtained distributions. A clear evolution as a function of time of the lags can be seen in  Extended Data Figure~\ref{fig:ccfs}-a .

We also checked the IR/UV lags. The HST lightcurve has an absolute timing accuracy up to $\approx 1$~s. We therefore computed the CCF between HST and HAWK-I to search for any possible lags with a time resolution of 1~s. As shown in Extended Data Figure~\ref{fig:ccfs}-b, the CCF peaks at 0. Some asymmetry is seen for lags of $\approx 50$~s, which can be explained by the different shapes of the flares. 

\underline{\textit{Fast flaring:}}

In order to study the properties of the fast flares, we selected spikes brighter than 0.7~mJy, finding 11 events in the lightcurve. We then selected 400-s segments centered on the peak values of each flare and averaged them (see Figure~\ref{fig:dataset}-d). Given the duration of each data cube of 50~s, we filled the readout gaps using a linear interpolation between the points at the edge, with the addition of Gaussian noise.  One flare (taking place around $\approx6300$~s from the start of the observations) showed two peaks separated by $\approx60$~s and was not included in the average. However, its inclusion does not affect the overall trend (Extended Data~Figure~\ref{fig:spikes}), confirming the connection between short and long timescales.

If these flares are produced by thermal gas with non-relativistic bulk motion, their typical timescale of $\approx 1$~s indicates a maximum size of the emitting region of $R \approx 10^{10}$~cm. In addition, their average IR flux density of approximately 1~mJy (which corresponds to $\approx 10^{-14}$~erg\,cm$^{-2}$\,s$^{-1}$), together with the source distance of 13~kpc\cite{buison_2020_1858_burst}, yields a brightness temperature of $\approx10^6$~K. Under these conditions, the thermal gas would emit an X-ray flux of $\approx10^{-8}$~erg\,cm$^{-2}$\,s$^{-1}$, which is roughly one order of magnitude higher than the observed one. This discrepancy cannot be solved assuming a smaller IR emitting region, as that would lead to a higher brightness temperature and X-ray flux. {Even considering effects of obscuration, with an NH of 10$^{23}$ cm$^{-2}$ and a covering fraction of 0.9 (i.e. significantly higher than the ones reported by previous X-ray studies\cite{hare2020}), would give an observed X-ray flux of at least a factor of a few higher.} We can therefore rule out a thermal origin for the flares. Given also their red color\cite{paice2018}, we interpret them as direct evidence of optically-thin synchrotron emission from transient/short-lived relativistic jet ejections expected  to take place\cite{nayakshin2000}$^{,}$\cite{janiuk2005} during these beat oscillations.

\subsection*{Reprocessing modelling}

We tested whether the evolving lags between X-ray and IR are compatible with reprocessing, with a previously used code simulating the expected delay between an X-ray flare from the neutron star and the corresponding peak in the IR due to the irradiation of the disk and the companion\cite{obrien,vincentelli2020_burst}.

For the binary we used the parameters of Swift J1858\cite{buisson2020_1858_eclipse}. In particular, we set the orbital period to 76841.3~s and tested inclinations between 70 and 90$^\circ$. We fixed the neutron star mass to 1.4 $M_\odot$ and used the same relation between the inclination and the mass ratio defined in past studies\cite{buisson2020_1858_eclipse}. For the companion temperature we adopted 6100~K (Castro Segura in prep.). Since it is better to have a sharp X-ray flare to measure the delay, we used a triangular function of width 20~s, with the peak at 10~s. We modelled the binary at a range of orbital phases from $\sim 20^\circ$ to $\sim 90^\circ$ in order to cover the 5 epochs at which we have measured lags. 

The delay between the X-ray and the IR peaks depends slightly on the shape we use for the X-ray flare, but we do find lags compatible with the observations for an inclination around $\sim80^\circ$. In general, we find that when the companion is not lying too close to the line between the observer and the neutron star, the lag is dominated by the IR reprocessing off the companion. This lag diminishes as the donor becomes more aligned with the observer-NS direction, until a minimum is reached. At that point the lag begins to increase again as the outer disk reprocessing dominates. The minimum lag allowed increases with inclination $i$, because the mass ratio changes with $i$. We fit our theoretical models to two sets of measured lags: the ones obtained including the flares in the lightcurves, and the ones excluding them (see left panels of Extended Data Figure~\ref{fig:reproc}, top and bottom respectively). The data points follow the theoretical trend remarkably well in both cases. When we include the flares, the posterior distribution is slightly skewed, preferring higher inclinations, with median at $i = 80.3^\circ$ (0.5, 16, 84, 99.5 percentiles at $76.2^\circ$, $78.4^\circ$, $82.0^\circ$, $84.1^\circ$). When we exclude the flares, due to the lower value of the lag at phase $\sim 45^\circ$ (which is also the least precise), the distribution prefers lower inclinations. At any rate, the result is compatible with the previous result, having median at $i = 79.3^\circ$ (percentiles at $75.6^\circ$, $77.6^\circ$, $81.1^\circ$, $83.5^\circ$). This figure confirms that the lag is indeed due to reprocessing of the X-rays. These values are fully consistent with the inclination obtained by modelling the eclipse profiles during the final stages of the outbursts.

\subsection*{Fast IR flare modelling }

A consistent interpretation of the fast IR flares needs to reconcile the following features: i) the IR flare duration time in the observer frame ($t_\mathrm{IR}$); ii) the observed frequency of the emission ($\nu_\mathrm{IR}$); iii) the peak flux density of the flare ($F_\mathrm{IR}$); and iv) that a bright flare is not simultaneously seen in the radio band. Here we introduce a simplified physical model that is capable to explain these features and derive additional information on the source from these constraints. In what follows we denote quantities in the jet reference frame with a prime. The relevant transformations between reference frames are given by $t = t'/\delta$, $\nu = \delta \nu'$, and $F = \delta^3 F'$, where $\delta(\Gamma, \theta)$ is the jet Doppler factor\cite{tetarenko2017}. For this calculation we will assume that the jet is launched perpendicularly to the accretion disk, such that $\theta = i$, and with semi-relativistic velocities, $\beta = 0.7 \pm 0.2$ (jet Lorentz factor $\Gamma_\mathrm{j} = 1.4 \pm 0.2$).

Both the synchrotron cooling time of a particle, $t_\mathrm{syn} \propto B^2 \, E_\mathrm{e}^{-1}$, and the characteristic frequency of the emitted photons, $\nu_\mathrm{syn} \propto B \, E_\mathrm{e}^2$, depend on the energy of the electron, $E_\mathrm{e}$, and the magnetic field intensity in the emitter, $B$ \cite{Blumenthal1970}. 
We assume that the duration of the flare is given by the electron synchrotron cooling time, that is, condition i) translates into $t'_\mathrm{syn}(E,B) = t'_\mathrm{IR}$. Moreover, the observed photon frequency should match the synchrotron frequency, so that condition ii) reads  $\nu_\mathrm{IR} = \nu_\mathrm{syn}$. Thus, from conditions i) and ii) we can derive the magnetic field intensity in the jet and the energy of the emitting electrons ($B \approx 553 \pm 22$~G and $E_\mathrm{e} \approx 207 \pm 8$~MeV, respectively). One caveat of this is that the magnetic field should not vary significantly during the duration of the flare in the jet frame ($\sim 6$~s).
In addition, one can assume that the observed flux is due to optically thin synchrotron emission from a relativistic electron population of the form $N(E) dE = N_0 E^{-p} dE$, with $p\sim2.2$ (see radio modelling section). We further assume that the minimum and maximum energies of the electrons are 1 MeV and 1 TeV, respectively, and normalize the electron distribution by adopting an equipartition energy condition, $U_\mathrm{e} = U_B$, which ties the energy densities in relativistic electrons and the magnetic field. With that, we can obtain $N_0$ from $U_\mathrm{e} = \int_{E_\mathrm{min}}^{E_\mathrm{max}} E N(E) dE$, and use it to calculate the optically-thin synchrotron SED at any given frequency \cite{Blumenthal1970}. Considering that the emitter is homogeneous and spherical, from condition iii) we can estimate its size to be $R_0 \approx 4.7\pm0.2 \times10^{10}$~cm. Thus, the IR-emitting jets are more compact and with much higher magnetic fields than the radio jets. 
Finally, we verify that condition iv) holds. Having the emitter size, the electron population, the magnetic field and the synchrotron luminosity, we can calculate the SSA opacity \cite{tetarenko2017}. We obtain $\tau \sim 10^6 \gg 1$ at 10 GHz, and $\tau \sim 10^{-6} \ll 1$ at IR frequencies, showing that the derived values are consistent with the proposed scenario.  We note that the overall conclusions do not change if we assume a broader range of jet velocities: adopting $\beta \ll 1$ leads to variations in the results below 25\%, whereas for $\beta = 0.99$ ($\Gamma_\mathrm{j} = 10$) variations are within a factor two, except for the flare time in the jet frame that increases up to $\sim 30$~s.

\subsection*{Radio modelling }
 
\underline{\textit{Swift J1858}}:
Jet synchrotron emission is known to produce strong radio variability, often in the form of flaring activity displaying a distinct observational signature where longer wavelength signals appear as a delayed, smoothed version of shorter wavelength signals. As our radio data exhibit this behavior, we attempted to model them and in turn quantify any relationship between the IR flares and the variations observed at longer wavelengths\cite{vanderlaan}. We implement the van der Laan synchrotron bubble model\cite{vanderlaan}, using the formalism outlined in Tetarenko et al.
\cite{tetarenko2017}. Briefly, this model describes the emission from ballistically moving, adiabatically expanding, bi-polar jet components, folding in projection and relativistic effects. To perform the modelling, we use a Markov Chain Monte Carlo algorithm (\textsc{emcee}\cite{Foreman-Mackey2013}), simultaneously fitting both radio frequency bands. Here we allow each ejected pair of jet components to have a different ejection time, peak flux, and speed, while tying the opening angle, inclination angle, magnetic field, and electron energy distribution (assumed to follow a power-law of the form $N(E) dE\propto E^{-p} dE$, where $p$ is related to the optical depth at the peak of the flare via $e^{\tau_p}-(2p/3 +1)\tau_p+1=0$) across all components. The radii of the ejecta at the peak of the flare are determined via $R_p=\left(\frac{F_p d^2}{\pi S_\nu(p,B_p)}\frac{1}{1-e^{-\tau_p}}\right)^{0.5}$, where $S_\nu(p,B_p)$ is the synchrotron source function dependent on the electron energy distribution and magnetic field at the peak of the flare. We use wide uniform priors for most model parameters, but sample from the known distance distribution ($12.8\pm0.18$~kpc, with bounds of 9--18 kpc), and use previous constraints on inclination in the literature to define this parameter's prior (uniform between 70--90$^\circ$\cite{buisson2020_1858_eclipse,buison_2020_1858_burst}). We ran two modelling cases; (1) ejection times are unconstrained (see Extended Data Figure~\ref{fig:model_unconstrain} and Extended Data Table~\ref{figt:1858_vdl_unconstrain}), (2) ejection times are constrained to only lie within the IR beats, as we may expect if both the IR and radio are synchrotron in origin (see Figure~\ref{fig:radio-modelling} and Extended Data Table~\ref{figt:1858_vdl_constrain}).

We find that both the modelling runs do a reasonable job of reproducing the radio light curves, capturing most of the flaring structure across both bands, with a similar quality of fit between runs. Our best-fit models contains 4 jet ejection events, where the ejection times coincide with IR flares occurring during the rapid variability periods. Although, the connection between radio and IR is more pronounced in modelling case (2), as we are better able to match up specific jet components to individual IR flares with lower uncertainty. Figure~\ref{fig:radio-modelling} displays the best-fit model for modelling case (2), overlaid on our radio light curves, and compares best-fit ejection times to the IR light curves (best-fit parameters are given in Extended Data Table~\ref{figt:1858_vdl_constrain}). Overall, we find that as we only need 4 ejections to reproduce the radio emission, most of the IR short flares do not actually become observable radio jet ejections. Although, we do notice that the jet ejection times seem to cluster near the boundaries of the rapid variability periods.
 
\underline{\textit{GRS 1915+105}}

With the GRS 1915+105 data, we follow the same modelling procedure as that of Swift J1858, allowing each ejected pair of jet components to have a different ejection time, peak flux, and speed, while tying the opening angle, inclination angle, magnetic field, and electron energy distribution across all components. We use wide uniform priors for most model parameters, but sample from the known distance ($8.6\pm2.0$ kpc \cite{redi2014}) and inclination angle ($60\pm5^\circ$ \cite{redi2014}) distributions. With GRS 1915+105 we do not place any constraints on ejection times. Here we use an incremental approach, building up our model flare by flare, to find the minimum number of components we need to reasonably reproduce the radio light curves. Additionally, as we find that the data does not allow us to put tight constraints on the electron energy distribution index, we fix this parameter to $p=2.2$ (equivalent to $\tau_p=1.6$, as expected from a Fermi acceleration mechanism via a single shock, commonly assumed for X-ray binary jets \cite{bland87,bell78,markoff2001}).

We find that in GRS 1915+105 we can reproduce the radio light curves reasonably well with 11 jet ejection events. Figure~\ref{fig:radio-modelling} displays the best-fit model overlaid on our radio light curves, and compares best-fit ejection times to the X-ray light curves (best-fit parameters are given in Extended Data Table~\ref{figt:1915_vdl}). {Upon comparing the best-fit parameters between Swift J1858 and GRS 1915, we find that the magnetic field strength of the ejecta are quite different between the two sources. This difference appears to be a natural consequence of the model. Given that the ejecta radii in the model are proportional to magnetic field strength and peak flux (among some other dependencies such as distance), to produce similar ejecta radii in both sources (as expected from the similar flare duration timescales) the lower peak fluxes in Swift J1858 with respect to GRS 1915 naturally result in a larger magnetic field estimate for Swift J1858. Moreover, similarly to what has been observed in past studies}\cite{mirabel1998}$^,$\cite{harmon1997}, the ejection times in GRS 1915 appear to lie in the dips between the rapid variability periods (seen here in the X-ray rather than the IR of Swift J1858). However, we notice that a sub-set of lower peak flux components have ejection times that lie close to the boundaries of the rapid variability periods. Interestingly, these fainter components seem to follow brighter ejections that tend to be launched closer to the beginning of a dip period rather than in the middle of it. 

To verify our key result in regards to the timing of the jet ejections in GRS 1915+105, and to evaluate how model dependent such a result was, we performed additional modeling runs under different conditions. In particular, we performed two other modeling runs using an 8 jet component model (i.e., without the extra lower peak flux components) with the electron energy distribution index fixed and free. The fixed model can reproduce the shoulders in the early flares but still has trouble reproducing the later peaks, while the free model can reproduce the flare decays well but not the shoulder features. Additionally, the free model favours a much lower electron energy distribution index of $p\sim0.8$, which lies outside the typical range ($p=2-3$) produced with the Fermi acceleration via a single shock mechanism commonly assumed in these types of jets. However, regardless of the differing model properties used in these extra runs, we find that all modeling runs still produce the same ejection time pattern, with the jet ejection times coinciding with the dip periods between variable sections of the X-ray light curve.

\subsection*{Timescale comparison }

The viscous timescale defines the time it takes for a mass accretion rate fluctuation to travel through the disc at a given radius. This can be expressed as\cite{belloni1997b}
\begin{equation}
    t_\mathrm{vis} \propto \alpha^{-1} ~M^{-\frac{1}{2}}~R^\frac{7}{2}~\dot{M}^{-2}, 
    \label{eq:one}
\end{equation}
where $\alpha$ is the viscosity parameter of Shakura and Sunyaev, $M$ is the mass of the compact object, $R$ is the radius of the disk and $\dot{M}$ is the absolute accretion rate.

Because the duration of the limit cycle is the viscous time in the unstable region of the disc, we find on average a similar $t_\mathrm{vis}$ for our BH and NS sources. Taking the ratio of Equation \ref{eq:one} for both sources and assuming similar values for $\alpha$ we get 
\begin{equation}
 \left(\frac{M_\mathrm{BH}}{M_\mathrm{NS}}\right)^{-1/2}~ \left(\frac{R_\mathrm{BH}}{R_\mathrm{NS}}\right)^{7/2}~ \left(\frac{\dot{M}_\mathrm{BH}}{\dot{M}_\mathrm{NS}}\right)^{-2} =1
 \label{eq:new}
\end{equation}
 
As both sources exhibit instabilities we assume that they are at the same accretion rate in Eddington units. Therefore  $\dot{M}_\mathrm{BH}/\dot{M}_\mathrm{NS}= M_\mathrm{BH}/{M}_\mathrm{NS}$. From these assumptions it follows that:
\begin{equation}
R_\mathrm{NS} =  R_\mathrm{BH}   \left(\frac{M_\mathrm{BH} }{ M_\mathrm{NS}}\right)^{-{5}/{7}}
\label{eq:final}
\end{equation}

Defining $r = R/R_\mathrm{S}$ where $R_\mathrm{S}= 2GM/c^2$, we find that, $r_\mathrm{NS} =  r_\mathrm{BH}  \left( \frac{M_\mathrm{BH}}{M_\mathrm{NS}} \right)^{2/7}$. Given a mass of 12.6~$M_\odot$ for GRS 1915+105 and 1.4~$M_\odot$ for Swift J1858, we have $R_\mathrm{NS} = 0.2 R_\mathrm{BH}$ or $r_\mathrm{NS} = 1.8  r_\mathrm{BH}$.

\subsection*{Swift J1858 O-IR observations and GRS 1915+105 X-ray heartbeats comparison:}

In order to quantify the association between the Swift J1858 O-IR beats and GRS 1915+105 X-ray heartbeats we compared the mean-normalized flux distribution of the \textit{Chandra}, HAWK-I, the Liverpool Telescope and optical time-series from two additional epochs.

{Optical data were obtained within 2019 in two epochs from the Hale 200-inch (5.1m) Telescope at Palomar observatory. In June 2019 Swift J0858 was observed using CHIMERA\cite{2016MNRAS.457.3036H}, a 2-band rapid photometer mounted on the Hale Telescope.
The instrument uses frame-transfer, electron-multiplying CCDs to achieve 15 ms dead time and is ideally used to study sub-second variability. 
Observations were obtained in g' and r' filters while CHIMERA was operated using the conventional amplifier with 0.2--1.0 second exposures with $2\times2$ binning. Images were bias subtracted and flat-field corrected while aperture photometry was carried out using the standard CHIMERA pipeline \cite{2016MNRAS.457.3036H}. Fluxes reported in this paper were estimated via differential photometry with respect to nearby stars.

In 2019 October 4 the field was observed by Palomar’s Wafer-Scale camera for Prime (WASP\cite{2017JATIS...3c6002N}) wide field prime focus camera using 10 s exposures, while the additional dead-time between observations is of the same order. 
Bias subtraction, flat field correction, and cosmic ray
removal were performed with IRAF\cite{1986SPIE..627..733T}. The night conditions were not photometric and differential aperture photometry was also performed in comparison to nearby stars.}

We divided the lightcurves in 20 flux bins and computed their cumulative distribution functions. As shown in Extended Data Figure~\ref{fig:cdf}, the only apparent difference is between the \textit{Chandra} and IR flux distributions at a mean-normalized flux of 0.7. However, due to the different amplitude between the X-ray and O/IR flares, we also computed the distribution of the X-ray after applying a linear rescaling to make the X-ray minimum and maximum match the O/IR. A Kolmogorov-Smirnov test indicates that the GRS 1915+105 X-ray lightcurve and the Swift J1858 OIR lightcurve are consistent with a single flux distribution. This analysis demonstrates that the O/IR variability patterns observed in August in Swift J1858 are commonly seen throughout the entire outburst and that they are compatible with the limit cycles in GRS 1915+105. 

\smallskip
 
\subsection*{{Flux-Flux diagram:}} 

We computed a flux-flux diagram using the time interval containing all bands from X-rays to IR (i.e. from 500~s to 2000~s).  We binned the time series to 10~s (i.e. the LT time resolution) and computed the correlation diagram between UV and the other bands. For the HAWK-I lightcurve, gaps were filled using a linear interpolation between the points at the edge, with the addition of Gaussian noise.  While the correlation between X-ray and UV is relatively poor ($\rho\approx0.4$), the correlation between UV, Optical and IR is clearly visible (see Extended Data  Figure~\ref{fig:flux-flux}-a). The powerlaw index of the correlations were $\beta_\mathrm{HAWK-I}=0.3\pm0.02$ and $\beta_\mathrm{LT}=0.43\pm0.03$ for HAWK-I and LT, respectively. This indicates stronger variations in bluer bands \cite{paice2018}, suggesting a thermal reprocessing scenario. This is confirmed when looking at constant IR to optical and the UV to optical ratios (i.e. the magnitude colour) for X-ray count rates $>1$ ct s$^{-1}$ (see Extended Data Figure~\ref{fig:flux-flux}-b).

\subsection*{{Comparison with other sources:}} 

Past studies indicated the presence of "GRS 1915 like" patterns in different neutron stars (most notably the Bursting Pulsar\cite{court2018}, the Rapid Burster\cite{bagnoli2015} and the  ULX NGC 3621\cite{motta2020}). Our multiwavelength analysis would therefore indicate that similar instabilities would take place also in these sources. As shown for the case of the Bursting Pulsar GRO J1744$-$28\cite{monkkonen}, the condition to show accretion disk instabilities is when the magnetospheric radius $R_m$ is larger than the star radius $R_*$. This means that the luminosity $L$ 
\begin{equation}
L > 2.75 \times 10^{38} \; \xi^{21/22} \; \alpha^{-1/11} \; m^{6/11} \; R_{*,6}^{7/11} \; B_{12}^{6/11} \; \mathrm{erg \, s^{-1}},
\label{equ:La_B}
\end{equation}
where $\xi$ is a factor between 0.5 and 1, $\alpha$ is the viscosity parameter, $m$ is the mass of the neutron star in units of 1.4 M$_\odot$, $R_{*,6}$ its radius in units of 10$^6$~cm and $B_{12} $ its magnetic field in units of 10$^{12}$~G. On the other hand, when the magnetospheric radius is equal to or smaller than the stellar radius the condition becomes:
\begin{equation}
L > 6.46 \times 10^{35} \; \alpha^{-1/8} \; m^{9/16} \; R_{*,6}^{5/16} \; \mathrm{erg \, s^{-1}}.
\label{equ:La_r}
\end{equation}
Finally, the magnetospheric radius is smaller than the star radius when
\begin{equation}
L > 2.83 \times 10^{45} \; \xi^{7/2} \; m^{1/2} \; R_{*,6}^{3/2} \; B_{12}^2 \; \mathrm{erg \, s^{-1}}.
\label{equ:rm_rs}
\end{equation}

According to these equations, depending on their magnetic field, neutron stars with a luminosity of at least $\approx10^{36}$ erg s$^{-1}$, may show a radiation pressure dominated inner disk (see Extended Data Figure \ref{fig:lvsb}). Therefore, we compiled a luminosity vs magnetic field diagram for a subset of sources to verify the self-consistency of our proposed scenario. As shown in Extended Data Figure \ref{fig:lvsb} all accreting neutron stars that show "GRS 1915-like" repetitive patterns lie on the radiation pressure dominated disk region of the parameter space \cite{bagnoli2015,motta2020}$^,$\cite{court2018}.  The plot also shows the typical ranges regarding standard classes of accreting neutron stars.

At low magnetic fields (<10$^9$ G),  we note that Swift J1858 lies  in the radiation pressure dominated area and, as indicated by recent X-ray/radio studies\cite{Rhodes2022}, is also between Atolls and Z sources. The latter group is known to have clear evidence of outflows and radio ejecta: this strongly suggests  that accretion instabilities may play a significant role in explaining the unique phenomenology of Z sources\cite{homan2016,motta2019}. Interestingly, we note that also the Rapid Burster lies in the radiation pressure dominated regime\cite{bagnoli2015}, even though the source does not always display such variability. Therefore, luminosity alone doesn't seem to be the only element necessary to trigger 1915-like patterns. 

As for intermediate magnetic fields between 10$^9$ and 10$^{11}$~G, the Bursting Pulsar\cite{court2017} is known to show strong repetitive burst patterns similar to GRS 1915+105\cite{court2017}. In this region of the parameter space Terzan 5 X-2 also showed outflows close to its peak luminosity \cite{miller2011} (i.e. within the radiation pressure regime), but non-stationary variability patterns were not observed.

 {For sources with a very high magnetic field (>10$^{12}$ G), the radiation pressure dominated disk solution requires a luminosity $L_X >5\times10^{38}$erg s$^{-1}$, i.e. almost in the Ultraluminous X-ray sources (ULXs) regime. For such high accretion rates it is argued that outflows are formed within the spherization radius\cite{2007MNRAS.377.1187P,2022MNRAS.509.1119M}. A particular interesting group of these sources are pulsating ULXs (PULXs), which due to their proximity and high brightness enable detailed studies of accretion in the super-Eddington regime. 
 Spectral and temporal studies within the last five years have allowed us to obtain the magnetic field, as well as evidence for the presence of outflows in several sources\cite{2018ApJ...857L...3W,2019MNRAS.488.5225V,2020MNRAS.491.4949V,2021A&A...651A..75F,2022ApJ...937..125B}, which define a clear region of the parameter space beyond the radiation pressure dominated disk transition.
Interestingly, we note that, as for the notorious Swift J0243.6+6124, some accreting pulsars can also behave as transient ULXs with luminosities well above the typical values of High mass X-ray binaries (HMXBs)\cite{vandeneijnden2018}. In fact, such systems have been seen already in the Magellanic Clouds\cite{2017A&A...605A..39T,2020MNRAS.494.5350V}. For these nearby transient ULXs and many of the traditional ULXs there is evidence of ionized emission and absorption that may be associated with outflows and disk winds \cite{2021MNRAS.508.3569K}.

The brightest source to display "GRS 1915-like" patterns is the ULX NGC 3621 \cite{motta2020}. However, the nature of its central object is still unknown. Therefore, even though the source is clearly in the radiation-pressure dominate disk region, it is not possible to fully constrain its position in the $L$ vs $B$ diagram.
 }

\makeatletter
\apptocmd{\thebibliography}{\global\c@NAT@ctr 34\relax}{}{}
\makeatother

\subsection*{Data availability Statement}
All raw data regarding the Swift J1858 August campaign and GRS 1915+105 \textit{Chandra}/\textit{VLA} observations are public and can be downloaded from their archive using the reported codes. All the reduced data from this campaign (including spectroscopic observations which are not presented here) will be made object of a publication and made accessible (Castro Segura et al. in prep.). Further analysis of the WASP and CHIMERA data is in progress, thus they are available on request to the authors.

\newpage

\section*{Extended Data}
 \renewcommand\figurename{Extended Data Table}

  \setcounter{figure}{0}

\begin{figure*}[!htb]
 \centering
 \caption{  \textbf{Swift J1858 jet modelling results without constraining the ejection time}. The third ejecta has values that significantly differ from the others (including very large errors, see also Extended Data Figure~\ref{fig:model_unconstrain}, t), suggesting the need of some additional kind of constraint.  We further report the radii (R$_p$) at 7.5 GHz based on the best-fit model parameters for each ejecta. Here the subscript ``$p$'' denotes values at the peak of the flare component at 7.5 GHz.} 
\includegraphics[width=0.8\columnwidth]{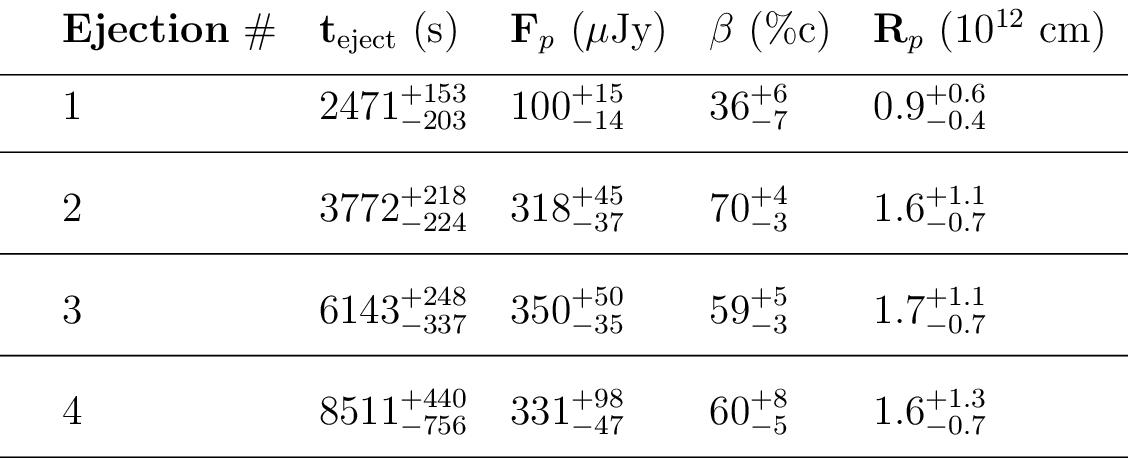}
\label{figt:1858_vdl_unconstrain}
\end{figure*}

\begin{figure*}[!htb]
  \centering
  \caption{ \textbf{Swift J1858 jet modelling results constraining the ejection time}. The 4 events were constrained to take place during the IR beats. Here the subscript ``$p$'' denotes values at the peak of the flare component at 7.5 GHz.   The model constrains also the quiescent 7.5 GHz radio level ($50_{-0.3}^{+0.1}\mu$Jy), the quiescent radio spectral index ($\alpha = -0.92\pm0.05$) , the opening angle ($\phi={3.3^\circ}_{-0.5}^{+0.1}$), the jet inclination ($i= {83.9^\circ}_{-0.4}^{+0.1}$), the optical depth at 7.5 GHz ($\tau_p=1.55_{-0.06}^{+0.04}$), and the magnetic field strength (B$_{p}=0.04\pm0.01$~G). The global values of the jet are consistent with the unconstrained fit. } 

\includegraphics[width=0.8\columnwidth]{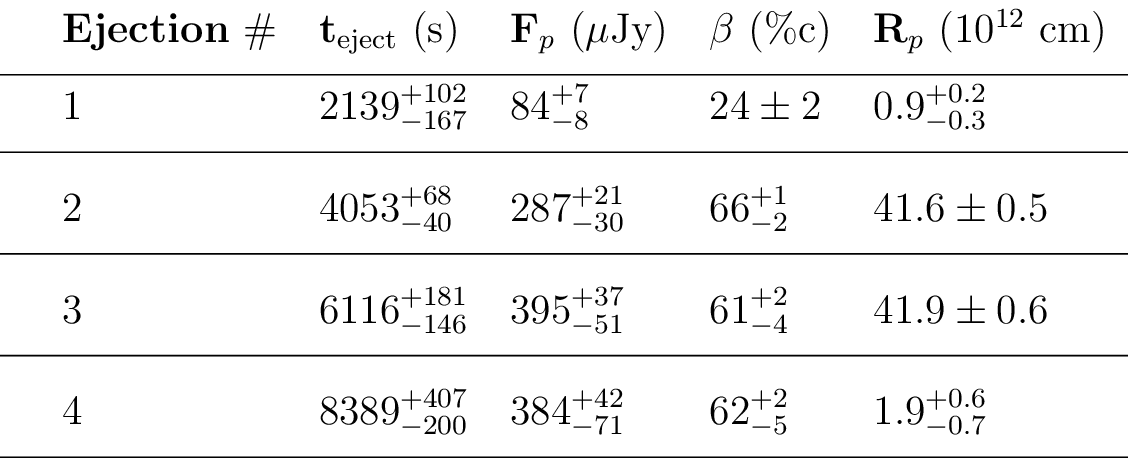}
\label{figt:1858_vdl_constrain}
\end{figure*}

\begin{figure*}[!htb]
 \centering
\caption{ \textbf{GRS 1915+105 jet modelling results}. We also obtained the best fit parameters for the quiescent 7.8 GHz radio level ($0.63\pm0.03$~mJy), the quiescent radio spectral index ($\alpha = -0.6\pm0.003$), the opening angle ($\phi={7.1^\circ}\pm0.04$), the jet inclination ($i= 62.7^\circ \pm0.03$), and the magnetic field strength (B$_{p}=0.05\pm0.002$~mG).  The optical depth at 7.8 GHz ($\tau_p$) was fixed to 1.6. We further report the component radii (R$_p$) at 7.8 GHz based on the best-fit model parameters for each ejecta. Here the subscript ``$p$'' denotes values at the peak of the flare component at 7.8 GHz.}
 
\includegraphics[width=0.8\columnwidth]{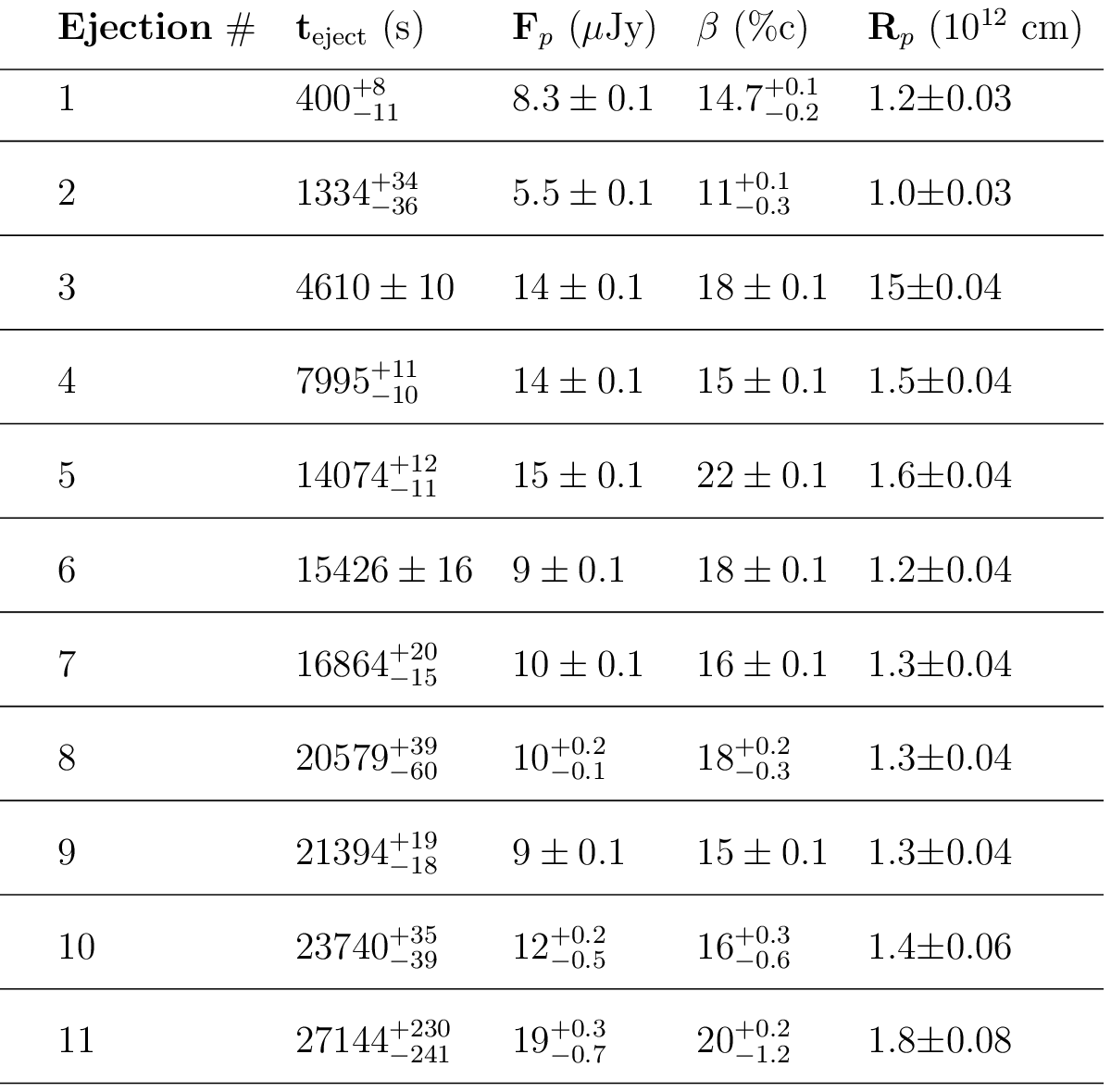}
\label{figt:1915_vdl}
\end{figure*}

 \renewcommand\figurename{Extended Data Figure}

\setcounter{figure}{0}

\begin{figure*} [!htb]
\centering
 \includegraphics[width=0.8\columnwidth]{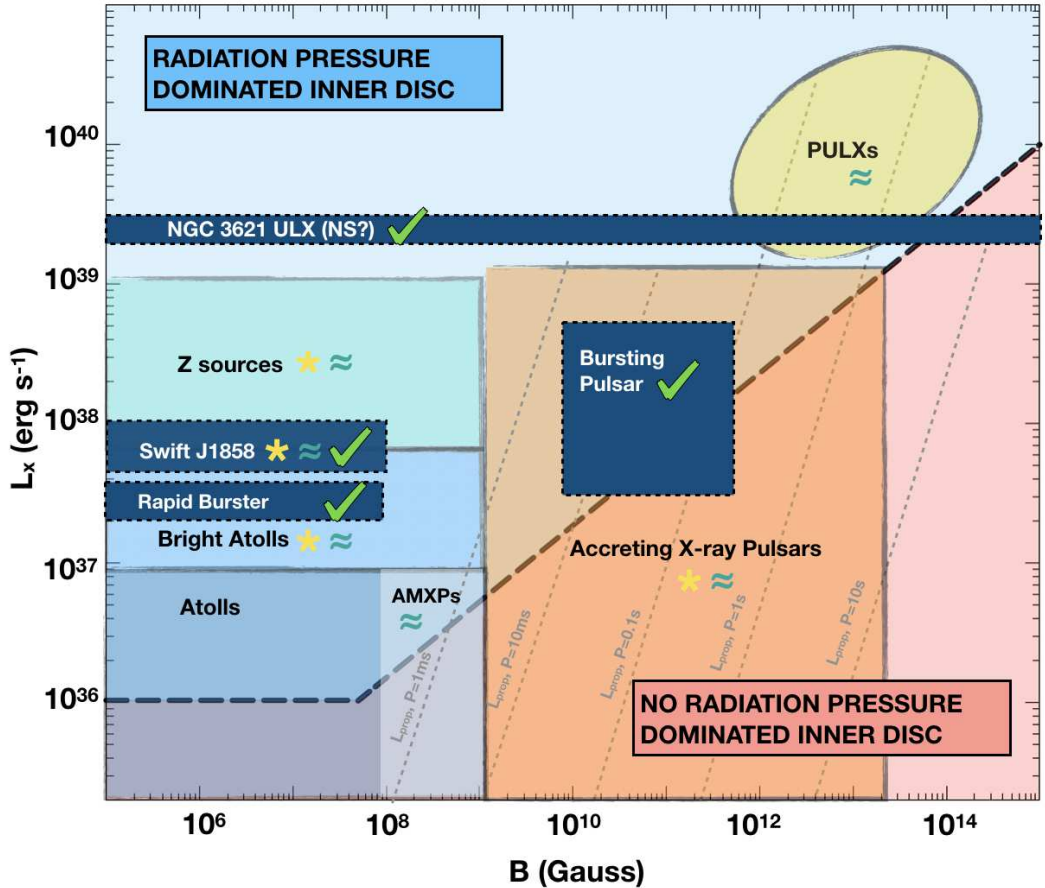}
\caption{ \textbf{Luminosity vs magnetic field diagram for accreting neutron stars. }
The long dashed line represents the limit in the parameter space for which the inner disk is (blue) or not (red) radiation pressure dominated: i.e. where the transition between zone A and B (R$_{AB}$) of a Shakura Sunayev\cite{shakurasunayev} disk is greater than the magnetospheric radius (R$_{m}$) or the star radius (R$_*$), whichever is larger.  
Since the disk rotates with Keplerian velocity, for large R$_{m}$ the inner disk rotates slower than the NS magnetosphere and accretion can be halted due to the propeller effect\cite{1975A&A....39..185I}.
The dashed grey lines correspond to the propeller threshold luminosities as a function  of the magnetic field for different spins\cite{1975A&A....39..185I,campana2002}. 
For all these lines we  set $\xi = 1$, $\alpha = 0.1$, M$_{NS}$ = 1.4 M$_\odot$ and R = 10 km. We then marked the sources which displayed "1915-like" variability patterns (green tick) in their lightcurves: the Rapid Burster\cite{bagnoli2015}, the Bursting Pulsar\cite{monkkonen,court2017}, the ULX NGC 3261\cite{motta2020} and our target Swift J1858. Along with these we also displayed the different phenomenological classes of accreting neutron stars X-ray binaries, depending on their magnetic field and luminosities.  At low magnetic field ($\leq$10$^9$~G) we find classical LMXBs, which, depending on their accretion rate, can manifest as Atolls, bright Atolls or Z sources \cite{homan2007,munioz-dariaz2014}. X-ray timing studies have also shown that accreting millisecond pulsars (AMXPs), with a magnetic field in the 10$^{8-9}$~G range, are also compatible with Atolls in their hard state\cite{gladstone2007}. 
For higher magnetic fields of 10$^{9}$~G, the observed accreting neutron stars usually show pulsations, but, due to the lower propeller threshold, have also slower spin periods with respect to AMXPs. Above 10$^{11}$~G, the diagram is mainly populated by High mass X-ray binaries\cite{coburn2022,reig2011} (HMXBs) 
and Pulsating Ultraluminous X-ray sources\cite{ulxreview}. For all these classes/objects we also marked which of them show typical phenomenology linked to accretion instabilities, i.e radio ejecta (yellow star) and outflows (cyan waves). We note that these phenomena tend to appear above the radiation pressure disk threshold  (see Methods). }%
\label{fig:lvsb}
\end{figure*}

\newpage

\begin{figure*}[!htb]
\centering
\includegraphics[width=\columnwidth]{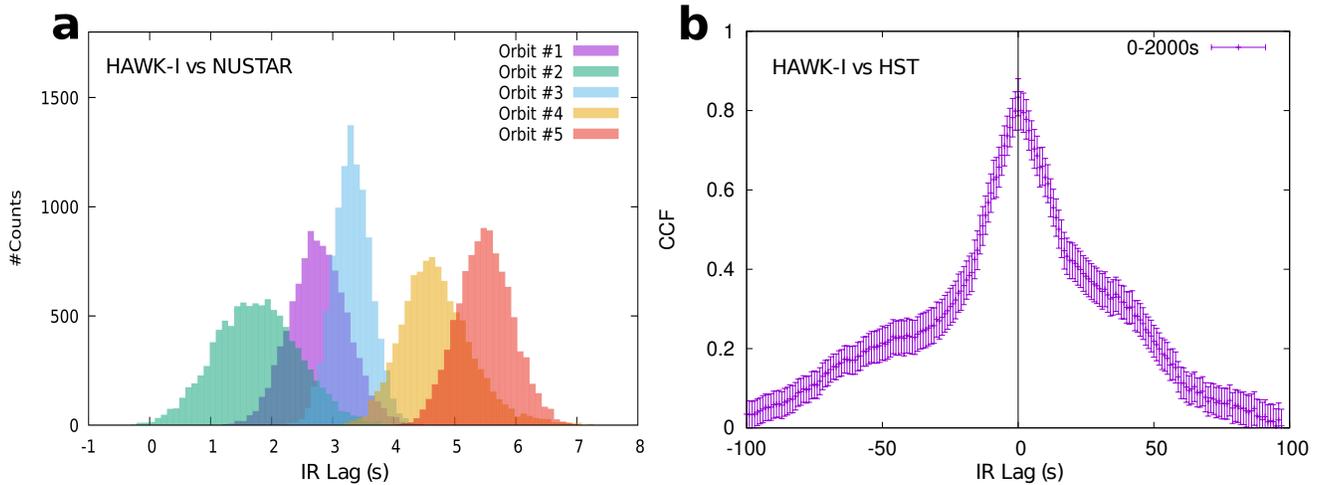}
\caption{ \textbf{Infrared Lag analysis}. \textit{(a)} Lag distribution from the 5 simultaneous \nustar--HAWK-I windows after 10$^4$ flux randomizations. An evolution of the lag centroid is visible. \textit{(b)} CCF computed between HST and HAWK-I. Excluding the asymmetry at longer lags, due to the asymmetry of the flares, the CCF peaks at 0.} 
\label{fig:ccfs}
\end{figure*}

\newpage
\begin{figure*}[!htb]
\centering
\includegraphics[width=0.7\columnwidth]{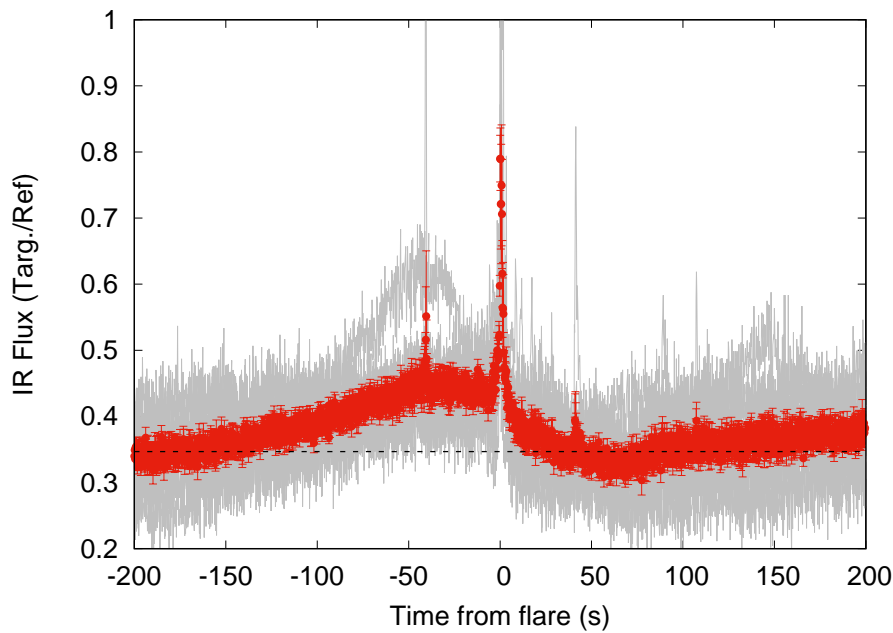}
\caption{\textbf{Averaged flare profile for events}. Due to the presence of two nearby flares, these appear also not at the center. However, the overall connection between long and short timescales is still clear.} 
\label{fig:spikes}
\end{figure*}

\newpage
\begin{figure}[!htb]
\centering  
\includegraphics[width=\columnwidth]{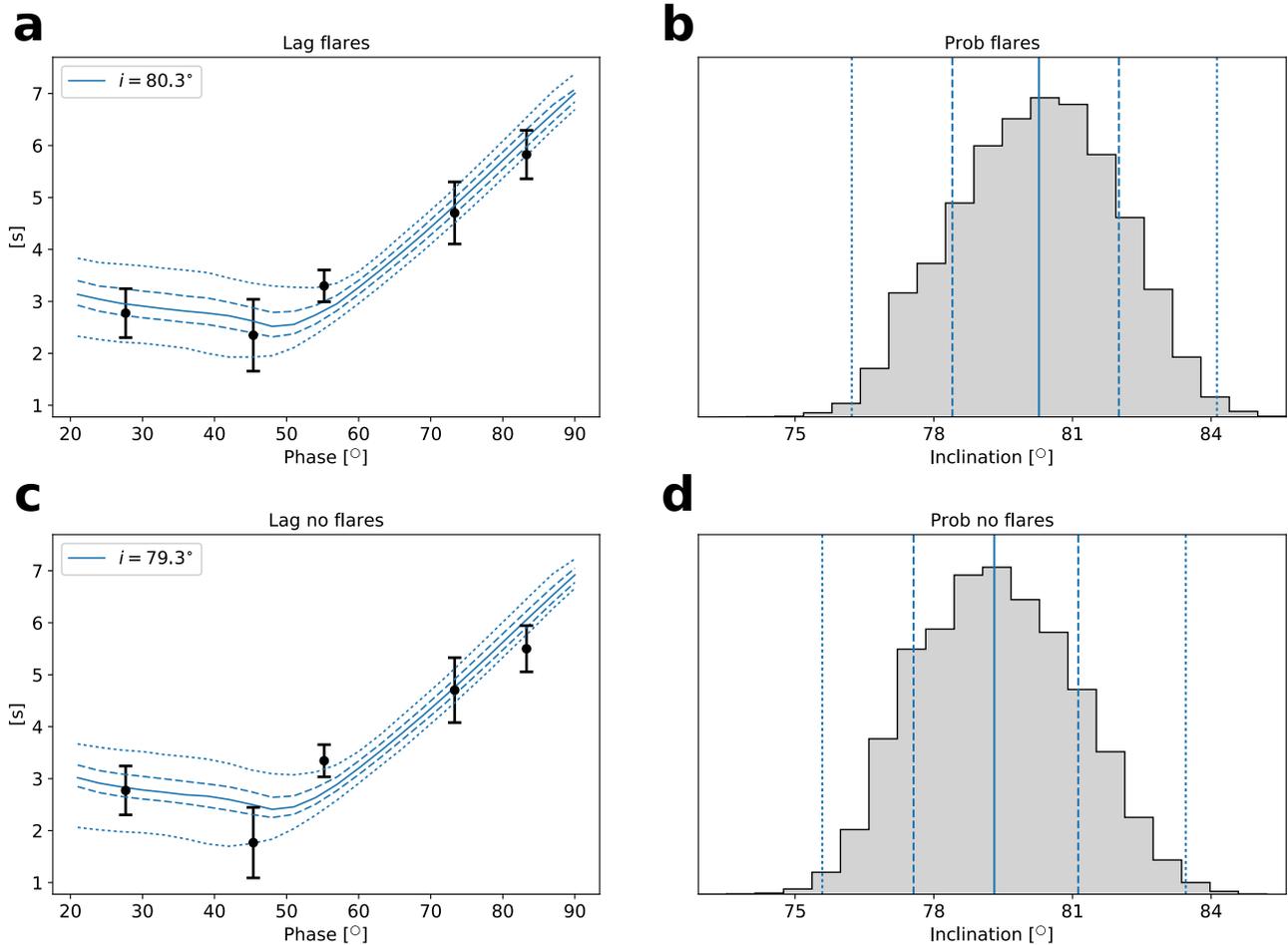}
\caption{\textbf{X-ray versus Infrared lag modelling}. \textit{(a)} Fit to the lags obtained including the flares in the lightcurves.  The only parameter allowed to change is the inclination of the binary, $i$. Long and short dashed curves represent 68\% and 99\% confidence level respectively. \textit{(b)} Histograms of the posterior distributions of the inclinations. In all plots dotted lines indicate the 0.5 and 99.5 percentiles, dashed lines indicate the 16 and 84 percentiles and solid lines indicate the median. \textit{(c)} Fit to the lags measured excluding the flares from the lightcurves. \textit{(d)} Histogram of the posterior distribution excluding the flares.}
\label{fig:reproc}
\end{figure}

\begin{figure*}[!htb]
\centering
\includegraphics[width=0.75\columnwidth]{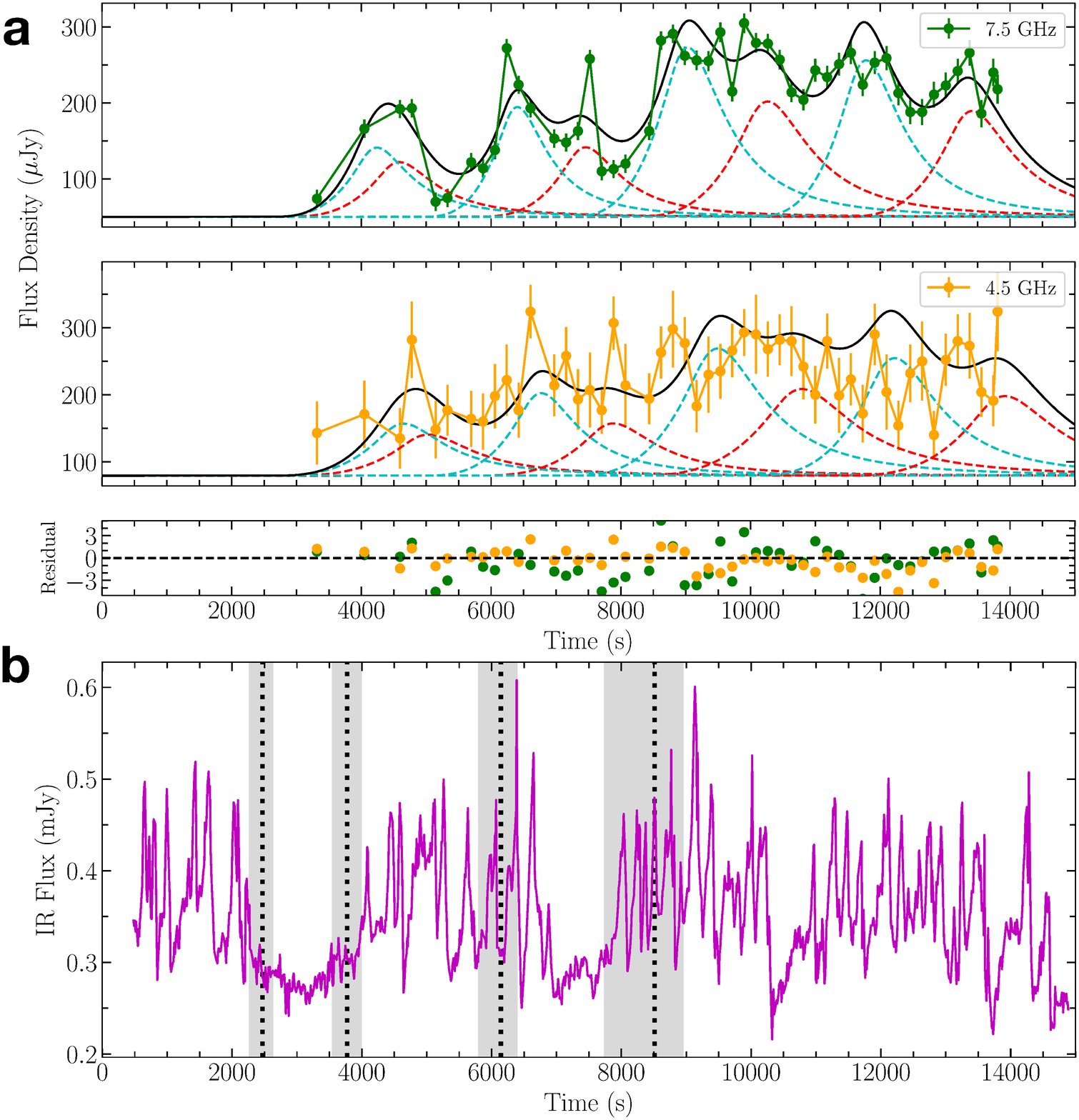}

 \caption{\textbf{Unconstrained radio modelling of Swift J1858}. Same as Figure~\ref{fig:radio-modelling}-a-c, but modelling the Swift J1858 radio light curve using no constraints on the ejection times. We found similar results with larger errors.} \label{fig:model_unconstrain}
\end{figure*}
\newpage

\begin{figure*}[!htb]
\centering
\includegraphics[width=\columnwidth]{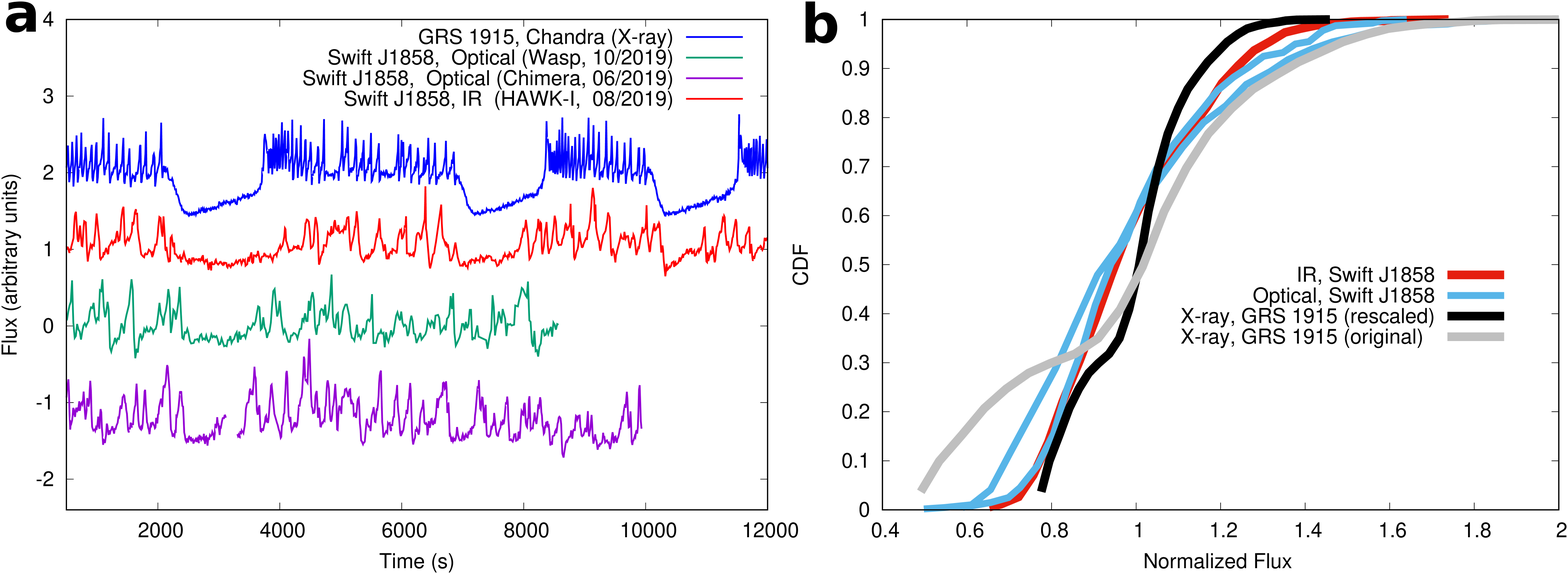}

\caption{ \textbf{Comparison of the beats at different wavelengths}. \textit{(a)} Averaged-normalized lightcurves used for quantifying the association between Swift J1858 and GRS 1915: \textit{Chandra} data from GRS 1915+105 (blue, +1 shift applied), HAWK-I data from Swift J1858 (red), WASP data from Swift J1858 (green, $-1$ shift applied) and CHIMERA data from Swift J1858 (purple, $-2$ shift applied). \textit{(b)} Cumulative distribution function (CDF) of the flux distributions of the lightcurves.} 
\label{fig:cdf}
\end{figure*}

\begin{figure*} [!htb]
\centering
 \includegraphics[width=\columnwidth]{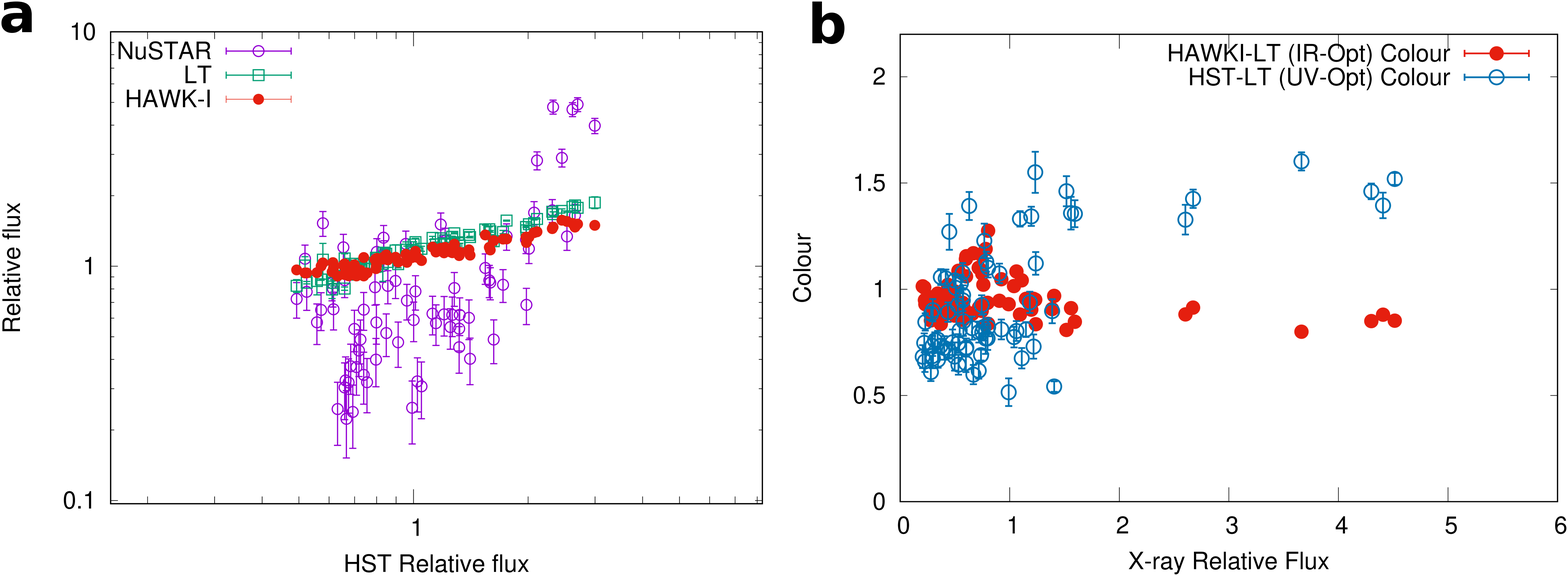}
  \caption{ \textbf{Flux-flux correlation diagram}. \textit{(a)} Flux-Flux diagram with respect to the HST measurements for \nustar ( open circles), LT (opened squares) and HAWK-I (filled circles). While the O-IR is well correlated, the X-rays show a non-linear trend. All bands have been normalized to their average. \textit{(b)} The plot shows the ratio of HAWK-I over LT (filled circles) and HST over LT (opened circles) as a function of the X-ray count rate normalized to 6.5 counts s$^{-1}$.}   \label{fig:flux-flux}
\end{figure*}

\end{document}